\documentclass[showpacs,preprintnumbers,amsmath,amssymb,twocolumn]{revtex4}

\usepackage{graphicx}
\usepackage{dcolumn}

\newcommand{\vb}[1]{{\mathbf{#1}}}
\newcommand{\lb}[1]{\label{#1}}

\newcommand{\bc}{\begin{center}}
\newcommand{\ec}{\end{center}}
\newcommand{\be}{\begin{equation}}
\newcommand{\ee}{\end{equation}}
\newcommand{\bea}{\begin{eqnarray}}
\newcommand{\eea}{\end{eqnarray}}
\newcommand{\ba}[1]{\begin{array}{#1}}
\newcommand{\ea}{\end{array}}

\newcommand{\bt}[1]{\begin{table}[ht]\centering\begin{tabular}{#1}}
\newcommand{\et}[1]{\end{tabular}\caption{\small#1}\end{table}}

\begin{document}

\preprint{SUSX--TH/02--006; hep-th/0202033} 

\title{From $U(1)$ Maxwell Chern-Simons to Azbel-Hofstadter:\\
Testing Magnetic Monopoles and Gravity to $\sim 10^{-15}$\textit{m}?}

\author{P. Castelo Ferreira\footnote{current address: CENTRA - Instituto Superior T\'ecnico, Av. Rovisco Pais 1, 1049-001 Lisboa, Portugal}}

\affiliation{Dep. de Mat. -- I. S. T., Av. Rovisco Pais, 1049-001 Lisboa, Portugal\\P. A. C. T. -- University of Sussex,
Falmer, Brighton BN1 9QJ, U.K.}
\email{pcastelo@catastropha.org}

\begin{abstract}
It is built a map between an Abelian Topological Quantum Field Theory,
$2+1D$ compact $U(1)$ gauge Maxwell Chern-Simons Theory
and the nonrelativistic quantum mechanics
Azbel-Hofstadter model of Bloch electrons. The $U_q(sl_2)$ quantum
group and the magnetic translations group of the Azbel-Hofstadter model 
correspond to discretized subgroups of $U(1)$ with linear gauge parameters.
The magnetic monopole confining and condensate phases in the Topological
Quantum Field Theory are identified with the extended (energy bands) and
localized (gaps) phases of the Bloch electron. The magnetic monopole condensate is
associated, at the nonrelativistic level, with gravitational white holes due
to deformed classical gauge fields. These gravitational
solutions render the existence of finite energy pure magnetic monopoles possible.
This mechanism constitutes a dynamical symmetry breaking which
regularizes the solutions on those localized
phases allowing physical solutions of the Shr\"odinger equation
which are chains of electron filaments connecting several monopole-white holes.
To test these results would be necessary a strong external
magnetic field $B\sim 5\ T$ at low temperature $T<1\ K$. To be accomplished, it
would test the existence of magnetic monopoles and classical gravity
to a scale of $\sim 10^{-15}$ \textit{meters}, the dimension
of the monopole-white hole. A proper discussion of such experiment is out of the scope of
this theoretical work.
\end{abstract}


\maketitle

\newpage

\addtolength{\baselineskip}{0.20\baselineskip}

\thispagestyle{empty}
\tableofcontents

\newpage

\setcounter{page}{1}

\setcounter{equation}{0}

\section{Discussion}

\subsection{Introduction and Results}

The motivations to this work are three fold.
First magnetic monopoles have not been observed experimentally.
Although due to the quantization of electrical charge, they should exist~\cite{Dirac}.
Secondly, testing gravity turn out to be a difficult task.
At large scales there is no assurance that black holes actually exist, although
very dense objects have been observed~\cite{bholes}. While at small scales it has only been
tested to the $\sim 10^{-3}$ \textit{meters} scale.
Thirdly several studies have address, from a theoretical point of view, the mechanisms
of confinement-condensate phase transitions in gauge theories.
Its relation with gravitation is usually neglected.

In this work is described such a confinement-condensate phase transition which
corresponds to a dynamical phase transition.
Gravitation plays an important role. Field configurations with
non trivial magnetic charge associated with metric deformations
makes the gauge theory in the magnetic condensate phase free of essential
singularities. These configurations constitute magnetic monopoles
which are also white holes~\cite{St_0,St_1,St_2,St_3,St_4}, they interact gravitationally
very weakly and reflect the incident matter. Here we call them monopole-white hole.

As far as the author knows the Localized phase for the Azbel-Hofstadter model are not considered in
the literature (except maybe for~\cite{PC_0,PC_1}), since the corresponding phase space region corresponds to the gaps of
the quantum theory. Although out of the scope of this work a possible experiment under strong magnetic fields
could in principle be attained as in~\cite{helium} where a confined phase for the electron was first observed.
In the framework presented in this work, and based in the results of~\cite{helium}, the conditions
for the existence of these configurations are in a regime of very low temperatures and
strong magnetic field. In order to test the theoretical results obtained it would be
necessary a strong magnetic field $B\sim 5\ T$ that would test the existence of magnetic
monopoles and test gravity to $\sim 10^{-15}$ \textit{meters} scale. This topic is clearly
not the expertise of the author and is not going to be developed here.

All the physical quantities on this work are expressed in SI units.

This work is organized as follows.
In Section~\ref{sec.TMGT} it is build a map between a Topological Massive Gauge
Theory~\cite{TMGT_01,TMGT_02,AHPS,BN_1,LR_1,AFC_1,TM_02,IK_2,TM_05,TM_06},
a Compact $U(1)$ Maxwell Chern-Simons Quantum Field Theory and the
Azbel-Hofstadter~\cite{H,brown,AZ_0,HO_0,KS,aa,wieg_1,wieg_2,wieg_3}
model of nonrelativistic electrons on a periodic potential under a strong
perpendicular magnetic field. The periodic potential in the former model
corresponds to a periodic current $J^\mu$ on the first theory which
reduces the gauge group to a discrete $U(1)$ (equivalently the quantum field theory
is defined on a lattice). It is shown that the discrete gauge
symmetries with linear gauge parameter correspond
either to the quantum group $U_q(sl_2)$ (note that this group is also connected with area preserving
diffeomorphisms~\cite{IK_1,IK_1a}) or magnetic translation group
depending on the gauge choice. The first correspond to
Landau gauge while the second to the symmetric gauge. Both
groups are in this way related to $U(1)$ trough the Topological Gauge Field Theory.

In Section~\ref{sec.AH} some results of the Azbel-Hofstadter model are analyse. It has two phases as studied
in~\cite{PC_0,PC_1}, one Extended, in which the magnetic
translations symmetry is present.
And the other one Localized, where that symmetry is broken. On compact
Maxwell Chern-Simons theories there exist vortexes
(or magnetic monopoles)~\cite{AK_1,AK_2,AK_3,AK_4} which may be in a
Confined or Condensate phase~\cite{AK_5,AK_6,AK_7}.
Given the map build in this work, the Confined phase of the gauge theory
corresponds to the Extended phase of the models, while the Condensate
phase to the Localized one. Here the terms condensation
and confinement are related to the condensation and confinement
of magnetic charge, meaning that magnetic charge can only be
physically observed on the condensate phase. In the Confinement phase only
magnetic dipoles are observed.

In Section~\ref{sec.gaps} the configurations on both phases are analysed.
In the Confined phase the gauge symmetry is present while in the Condensate phase the gauge
symmetry is broken. This phase transition is shown to
correspond to dynamical symmetry breaking~\cite{thooft_1,thooft_2,polyakov_1,polyakov_2,thooft_3}.
The symmetry breaking is achieved by considering a particular magnetic monopole solution of the gauge
fields together with a nontrivial metric configuration which
constitutes the white hole~\cite{St_0,St_1,St_2}. One crucial
consequence of this construction is that the association of
the white hole to pure magnetic monopoles renders the
energy of the configuration finite. In the Confined phase, gravity plays no special role, we
are in a quantum regime where the solutions of the Shr\"odinger equation are the usual Landau levels
in the presence of a magnetic field while in the Condensate phase we obtain electron filaments~\cite{helium}
attaching several magnetic monopole-white holes which we call in this work chains. We refer
the reader to the work~\cite{Dunne2} for $SU(2)$ BPS monopole chains.

\subsection{Open Issues}

The most important point to further explore seems to be the
relation between the gravity and gauge sectors of the theory.
The configurations presented in this work are given at ansatz level
without a full theoretical justification.
A semi-classical treatment is given in~\cite{St_1,St_2}.
However the solutions of these works do not possess similar
characteristics to our ansatze, namely the charge and energy of the configurations are not finite.
Also following~\cite{IK_3,CM_01,CM_02} the author studied classical solutions for gauge Chern-Simons and a
gravitational scalar field~\cite{PC_BH_01,PC_BH_02}. However the results
are not very promising. For the magnetic case~\cite{PC_BH_03}, although horizons do exist, the usual Maxwell
charges diverge. Only the Page charges are finite~\cite{charges}.

It is also important to stress that the usual $2+1$ gravity has no degrees of freedom. This
problem may be solved by considering noncommutative gravity~\cite{NCG_01,NCG_02,NCG_03,NCG_04}.
Although have been suggested that gravity may behave noncommutatively at short distances~\cite{NCG_01}
this subject is not physically well established. The problem is also linked to
the proper definition of gauge invariant coordinate transformations~\cite{J_01} in
noncommutative spaces (see~\cite{J_02} and references therein) and the constraints between
the gauge and the gravitational fields~\cite{madore}. Most approaches consist of considering
a first order correction to the commutative metric~\cite{NCG_03,St_2}, but it would be interesting
to study the solutions of the EOM corresponding to a full noncommutative metric.
The main draw back is that, in this case, we are dealing with complex gravity.

The values of temperature ($<1K$) and magnitude
of the magnetic field (5T) to achieve a condensate phase (Localized phase), are based on
the experimental results of~\cite{helium} and on the usual definition of strong magnetic
field. Nevertheless we did not study the precise values for the phase transition.
Based on the Coulomb Gas approach, Kogan and Kovner~\cite{AK_5} derived that
the phase transition corresponds to the value of the CS coefficient $1/\pi$ (in natural units $\hbar=1$).
In the framework presented here it corresponds to a relatively weak magnetic field $B=0.07435\ T$. The
conditions for the validity of the Coulomb Gas approach may not be appropriated for
a low matter density experiment in the laboratory but certainly it
could be appropriate for high matter density regions of our universe.

Once we are talking about magnetic charges, it would not be polite, not to discuss electromagnetic duality
(see~\cite{olive} and references therein).
Taking the usual quantization relation $eg=2\pi\hbar\,N$, between pure magnetic ($g$) and electric charges ($e$) we obtain
from equation~(\ref{Qm}) that $Q_{mon}:=B/4=2\pi\hbar N$. This simply implies
that the system will \textit{react} to values of the magnetic field when it raises (or decreases)
over some threshold $B=8\pi\hbar N$.

Also it remains to explain what are the dynamics
of the mechanisms presented here, namely how the
white holes emerge dynamical, is it a mechanism similar to the
formation of black holes~\cite{collapse} or is it
a completely different mechanism?

Finally the relation with $U_q(sl_2)$ and area preserving
diffeomorphisms~\cite{IK_1a} suggests some relation
with the Quantum Hall Effect. Note however that the
solutions presented here are in the Gaps of the
nonrelativistic quantum theory. They are not related
to the Landau level solutions for the energy Bands.
They correspond to Localized phases which (so far as the
author knows) have only been detected recently
on the experiments described and conducted by the authors
of reference~\cite{helium}.

\setcounter{equation}{0}
\section{\lb{sec.TMGT}From $3D$ Topological Massive Gauge Theories to Electrons
on Magnetic Fields}

In this section is considered a compact $U(1)$
topological massive gauge theory, Abelian Maxwell
Chern-Simons. This theory will have as a particular
limit an effective description of some other system,
electrons moving on perpendicular magnetic fields.
The coupling of the topological term $k$
will be interpreted as an effective
physical quantity, the external magnetic field.

\subsection{TMGT, $3D$ Maxwell Chern-Simons}

Let us start with a $3D$ Abelian Compact Maxwell Chern-Simons Gauge Theory,
\bea
S_{TMGT}&=&\displaystyle\frac{\mu_0\ \alpha_T}{\ V_{3}}\int_M d^3x\sqrt{-g}\left[-\frac{1}{4\gamma^2}F_{\mu\nu}F^{\mu\nu}+\right.\nonumber\\
& &\displaystyle\left.\frac{k}{8\pi}\epsilon^{\mu\nu\lambda}A_\mu\partial_\nu A_\lambda+A_\mu J^\mu\right]
\lb{STMGT}
\eea
where $F=dA$ is the usual connection of the
gauge field $A$ and a generic external $3D$
current $J^\mu$ for the gauge field was introduced.
The coupling constants are introduced on a nonstandard
way in order to simplify the phenomenology of the
nonrelativistic limit studied in this work.
$M$ is some Minkowski manifold allowing a $2+1$ splitting and
having a topology $M=\Sigma\times R$,
where $\Sigma$ is a $2D$ compact Riemann surface
(with $\partial\Sigma=0$) and  $R$ some finite interval which is going
to play the role of time, the metric is considered to be $2+1$
decomposable (see for example~\cite{AFC_1})
\be
ds^2=-dt^2+h_{ij}dx^idx^j
\ee
and therefore the factor containing the determinant of the metric reads
$\sqrt{-g}=\sqrt{h}$.
The antisymmetric tensor $\epsilon^{\mu\nu\lambda}$ is defined covariantly as
\be
\epsilon^{\mu\nu\lambda}=\frac{\hat{\epsilon}^{\mu\nu\lambda}}{\sqrt{h}}\ ,
\ee
where $\hat{\epsilon}^{\mu\nu\lambda}$ is the usual numerical antisymmetric
matrix with $\hat{\epsilon}^{012}=1$.

$V_{3}$ stands for a unit $3D$ spatial volume and has units of length cube ($m^3$),
$\alpha_T$ stands for an effective constant which relates the $2+1D$
physical system to the $3+1D$ physics.
In simple terms is related to the thickness of the system on the
perpendicular direction, say $x^3=z$, it has therefore units of length ($m$). The
physical meaning of $\gamma$ and $k$ will be
addressed later on.

These theories are usually named Topological Massive Gauge Theories
(TMGT)~\cite{TMGT_01,TMGT_02},
topological due to the topological character of the CS term (it is a
topological invariant) and massive since the gauge boson $A$ acquires a
topological mass due to the coexistence of both Maxwell and CS terms $M=k\gamma^2/4\pi$.
This can be checked by the equations of motion for $A$. Taking its curl
and rewriting it in terms of the dual field strength 
$(*F)^\mu=\epsilon^{\mu\nu\lambda}F_{\nu\lambda}/2$ we obtain $(\partial^2-M^2)(*F)^\mu=0$.

The Hamiltonian of such a system is easily computed to be
(for further details see for example~\cite{AFC_1,TM_05,TM_06}),
\begin{widetext}
\be
\ba{rl}
H_{TMGT}=&\displaystyle\int_\Sigma d^2x\sqrt{h}\left[-A_0\left(\partial_i\pi^i+\frac{k}{8\pi}\epsilon^{ij}\partial_iA_j-J^0\right)\right]+\\[4mm]
&\displaystyle\frac{\mu_0\ \alpha_T}{2\ V_{3}}\int_\Sigma d^2x\sqrt{h}\left[\frac{1}{4\gamma^2}(\epsilon_{ij}F^{ij})^2+c^2\gamma^2\left(\pi_i-\frac{k}{8\pi}\epsilon_i^{\ k}A_k\right)\left(\pi^i-\frac{k}{8\pi}\epsilon^{il}A_l\right)-A_iJ^i\right]\ ,
\ea
\lb{HTMGT}
\ee
\end{widetext}
where $\pi^i$ are the canonical momenta conjugated to $A_i$
\be
\pi^i=-\frac{1}{c^2\gamma^2}F^{0i}+\frac{k}{8\pi}\epsilon^{ij}A_j
\ee
and the $A_0$ was taken to be a Lagrange multiplier imposing the gauss law
\be
\partial_iE^{i}+\frac{k}{4\pi}b=J_0\ .
\ee
$J_0$ is some external charge distribution and
the \textit{electric} and \textit{magnetic} fields are defined as
\be
\ba{rcl}
E^i&=&\displaystyle\pi^i-\frac{k}{8\pi}\epsilon^{ij}A_j\\[3mm]
&=&\displaystyle-\frac{1}{c^2\gamma^2}F^{0i}\ ,\\[5mm]
b&=&\displaystyle\frac{1}{2}\epsilon^{ij}F_{ij}\\[3mm]
&=&\displaystyle F_{12}\ .
\ea
\ee
Note that the $2D$ antisymmetric tensor is actually induced from the
$2+1D$ dimensional one as $\epsilon^{ij}=\epsilon^{0ij}$.

Upon quantization the commutation relations are
\be
\left[\pi^i(\vb{x'}),A^j(\vb{x'})\right]=-i\,\hbar\,\delta^{ij}\,\delta^{(2)}(\vb{x}-\vb{x'})
\lb{ComAP}
\ee
and the commutation relations between the fields are computed
to be
\be
\ba{rcl}
\displaystyle\left[E^i(\vb{x}),E^j(\vb{x}')\right]&=&\displaystyle-i\,\frac{\hbar}{\mu_0}\,\frac{k}{4\pi}
\epsilon^{ij}\,\delta^{(2)}(\vb{x}-\vb{x}')\ ,\\[4mm]
\displaystyle\left[E^i(\vb{x}),b(\vb{x}')\right]&=&\displaystyle-i\,\frac{\hbar}{\mu_0}\,\epsilon^{ij}\,\partial_j\delta^{(2)}(\vb{x}-\vb{x}')\ ,
\ea
\lb{com_EB}
\ee
where we assume the $2d$ delta functions to be normalized with a unit factor of area $[V_2]=m^2$.

The Hamiltonian (in the absence of external
sources) depends only on the coordinate combinations
$\pi^i-k\epsilon^{ij}A_j/8\pi$ such that
\be
H=H\left(\pi^1-\frac{k}{8\pi}A_2,\pi^2+\frac{k}{8\pi}A_1\right)\ .
\ee
Being so, the most generic operator $O$ which commutes with
the Hamiltonian, must depend on the coordinate combinations
$\pi^i+k\epsilon^{ij}A_j/8\pi$ such that
\bea
\displaystyle\left[H\left(\pi^1-\frac{k}{8\pi}A_2,\pi^2+\frac{k}{8\pi}A_1\right),\ \ \ \ \ \ \ \right.&\\[3mm]
\displaystyle\left.O\left(\pi^1+\frac{k}{8\pi}A_2,\pi^2-\frac{k}{8\pi}A_1\right)\right]&=0\nonumber
\eea
This fact is simply proved by using the commutation
relations~(\ref{ComAP}).

The most usual and known example of such an operator is, of course,
the gauge transformation operator
\be
U_\Lambda=\exp\left\{i\,\frac{\mu_0}{\hbar}\frac{\alpha_T}{V_3} \int\limits_\Sigma\sqrt{h}\,\Lambda(\vb{x})\left(\partial_iE^i+\frac{k}{4\pi}b-J_0\right)\right\}\ .
\lb{U}
\ee
In this case the generator is simply a representation of $U(1)$.
Note that, accordingly, the gauss law $\partial_i\left(\pi^i+k\epsilon^{ij}A_j/8\pi\right)=0$ depends
only on $\pi^i+k\epsilon^{ij}A_j/8\pi$. We consider from now on that
the measure in the 2-dimensional integral is defined covariantly. For
the sake of compactness of the equations we do not write the area factor either,
so the factor $\sqrt{h}\,\alpha_T/V_3$ is implicit on all the $2d$ integration over $\Sigma$.

\subsection{Compact TMGT}

Being compact means that large gauge transformations are allowed.
That is that the fields are defined up to close shifts around the
holonomic cycles of the manifold $\Sigma$. The generators of this cycles
are none other than the Wilson lines
\be
W_\alpha=\exp\left\{i\oint_\alpha A_idx^i\right\}
\ee
for some non contractible close cycle $\alpha$. Also note that
these objects are gauge invariant and constitute a basis for the
topological wave functions of the theory as extensively
studied~\cite{BN_1,LR_1} (see also~\cite{TM_02,TM_06}
for more generic geometries).

Therefore $\Lambda$ is a
compact parameter, say in the interval $[0,2\pi]$
and the boundary conditions on the gauge fields
need to be compatible with the identification
\be
\Lambda\cong\Lambda+2\pi\ .
\ee

Using the Schr\"{o}dinger picture $\pi=i\delta/\delta A$,
the gauge transformations acting on wave functions has the effect
\be
U(\Lambda)\Psi[A]=\exp\left\{-i\,\frac{\mu_0}{\hbar}\int_\Sigma\Lambda\left(\frac{k}{8\pi}b-J_0\right)\right\}\Psi[A+d\Lambda]
\lb{UPsi}
\ee
which must be invariant under a shift $\Lambda\to\Lambda+2\pi$
due to the compactness of the gauge group.
Then it imposes the group (dimensionless) charge quantization condition
\be
q=s_1+\frac{\mu_0}{\hbar}\,\frac{k}{8\pi}\int_\Sigma b=s_1+\frac{\mu_0}{\hbar}\,\frac{k}{4}s_2\ ,
\lb{q}
\ee
for some integer $s_1$ and $s_2$. Note that $s_1$ stands for the winding
of the gauge group, i.e. how many times the gauge field winds on $\Sigma$
when a shift $\Lambda\to \Lambda+2\pi$ is considered.

In the case of the compact gauge theory there are local
operators which create magnetic vortexes
creating a charge of flux $2\pi N$ for some integer
$N$~\cite{AK_1,AK_2,AK_3,AK_4}. The physical Hilbert space
of the compact theory is obtained by constraining the
Hilbert space of the noncompact theory. This is done by demanding
invariance of the physical states under the action of any combination
(product) of those vortex operators. They are obtained
integrating by parts $U$~(\ref{U}) 
\be
V(\Lambda)=\exp\left\{i\,\frac{\mu_0}{\hbar}\int_\Sigma\left[\partial_i\Lambda\left(\pi^i+\frac{k}{8\pi}\epsilon^{ij}A_j\right)-J_0\Lambda\right]\right\}\ .
\lb{UV}
\ee

Interpreting $\Lambda$ as an angle in the $2D$ plane and using
the identity
$\partial_i\Lambda(\vb{x})=-\epsilon_i^{\ j}\partial_j\ln|\vb{x}|$ (from the
the Cauchy-Riemann equations) one obtains
\bea
V(\vb{x}_0)=\exp\left\{-i\,\frac{\mu_0}{\hbar}\int_\Sigma\left[\left(E^i+\frac{k}{4\pi}\epsilon^{ij}A_j\right)\right.\right.\times\nonumber\\
\displaystyle\epsilon_i^{\ k}\partial_k \ln |\vb{x}-\vb{x}_0|-\Lambda(\vb{x}-\vb{x}_0)J_0\Bigl]\Bigl\}\ .\hspace{10mm}
\lb{V}
\eea

This operator creates a vortex that generates
magnetic flux.
its commutator with the magnetic field is
\be
\left[b(\vb{x}),V^N(\vb{x}_0)\right]=2\pi\,N\,V^N(\vb{x}_0)\,\delta^{(2)}(\vb{x}-\vb{x}_0)\ ,
\ee
where the $2D$ identity $\partial^2\ln|\vb{x}|=2\pi\delta(\vb{x})$ was used.

For compactness of the equations, in the following subsection, we use the notation
$\Delta\vb{x}_0=\vb{x}-\vb{x}_0$.

\subsection{Vortex, Monopoles and Field Decomposition}

These operators can be also interpreted as generators
of discontinuities or cuts from $0$ to $\infty$ in the
complex plane that generate a change on
the charge (or tunnelling effect between states with the
same energy and different charges). On other words they are associated
with singular gauge transformations or
monopole-instantons processes.

Note that two operators $V(\vb{x})$ and $V(\vb{x}')$ do not
commute, generally
\bea
V(\Lambda,\vb{x}_0)\,V(\Lambda',\vb{x}_0')=\hspace{40mm}\lb{VVcom}\\[2mm]
\exp\left\{-i\frac{k}{4\pi}\frac{\mu_0}{\hbar}\int_\Sigma\partial\Lambda(\Delta\vb{x}_0)\times\partial\Lambda'(\Delta\vb{x}_0')\right\}V(\vb{x}_0')V(\vb{x}_0)\nonumber
\eea
where the integral
\bea
\displaystyle\int_\Sigma\int_\Sigma\epsilon^{ij}\partial_i\Lambda(\Delta\vb{x}_0)\partial_j\Lambda'(\Delta\vb{x}_0')\delta^{(2)}(\Delta\vb{x}_0-\Delta\vb{x}_0')=\nonumber\\[2mm]
\displaystyle\int_\Sigma\epsilon^{ij}\partial_i\Lambda(\Delta\vb{x}_0)\partial_j\Lambda'(\Delta\vb{x}_0)\hspace{30mm}
\eea
was considered and the explicit dependence on $\Lambda$ restored.

Also they have the property
\bea
V(\Lambda,\vb{x}_0)V(\Lambda',\vb{x}_0')=\hspace{40mm}\\[2mm]
\displaystyle\exp\left\{-i\frac{k}{2\pi}\frac{\mu_0}{\hbar}\int_\Sigma\partial\Lambda(\Delta\vb{x}_0)\times\partial\Lambda'(\Delta\vb{x}_0')\right\} V(\Lambda+\Lambda',\vb{x}_0)\ .\nonumber
\eea

So this operators indeed form a group, in this case is
simply $U(1)$ as we already know.

For future use consider the decomposition of the spatial
part of the gauge field $A_i$ into its longitudinal
$\vb{a}_1=\{a_{1\,i}\}$ and transverse $\vb{a}_2=\{a_{2\,i}\}$ components
\be
\vb{A}=\vb{a}_1+\vb{a}_2
\lb{A}
\ee
such that by definition
\be
\ba{ccc}
\ba{l}\left\{\ba{l}\nabla.\vb{a}_1\neq0\\[2mm]\nabla\times \vb{a}_1= 0\ea\right.\\[8mm]
      \left\{\ba{l}\nabla.\vb{a}_2= 0\\[2mm]\nabla\times \vb{a}_2\neq0\ea\right.
\ea&
\Rightarrow&
\left\{\ba{l}\nabla.A=\nabla.\vb{a}_1\\[4mm]\nabla\times A=\nabla\times \vb{a}_2\ea\right.
\ea
\lb{Adef}
\ee

With this decomposition the magnetic field $b$ depends
uniquely on the field $\vb{a}_2$. From the perspective
of nontrivial magnetic configurations this fact can
also be interpreted as $\vb{a}_2$ containing the
\textit{singular} part of the gauge field and
$\vb{a}_1$ the regular part. Here \textit{singular}
means that it includes the discontinuities (singularities)
that generate the magnetic charge.

Note that once we introduce an external source, it will
\textit{reduce} the symmetries of the theory. Nevertheless,
depending on the specific configurations it will
maintain some subgroup of the full gauge group.
We will use this fact in the following section.
A geometrical interpretation is that a
\textit{local} charge insertion (or more generally a current
insertion) may be interpreted as a vertex insertion on the manifold
$\Sigma$, this will clearly change the topology of the manifold, in
particular its holonomy group. Note that by each insertion we create a
new non contractible loop around that insertion.
Due to the massiveness of the photon, its \textit{perturbative}
effect decays exponentially although the topology has
effectively being modified. This matter
will reduce the gauge symmetry group of the theory since
physically it is creating an effective potential.

In the case of a $xy$ periodic lattice the $U(1)$ gauge
symmetry will be reduced to some discrete subgroup
of $U(1)$ as it will be argued below.

\subsection{Electrons on a Magnetic Field}

Consider a particular longitudinal gauge
field $\vb{a}_1=a$, such that 
\be
\ba{rcl}
A_i&=&4\pi \epsilon_i^{\ j}a_j\\[4mm]
1&=&|\nabla\times a|\\[4mm]
0&=&\nabla.a\ .
\ea
\ee
The second condition is simply a
choice of normalisation for the field $a$ and will be shown to be
compatible with both the Landau and symmetric gauge. As argued
above the magnetic field $b$ is null for longitudinal fields
\be
b=\frac{1}{2}\epsilon_{ij}F^{ij}=F_{12}=\partial_ia^i=0
\ee

Rewrite now~(\ref{HTMGT}) as the \textit{simplified}
Hamiltonian
\be
H_a=\frac{c^2\gamma^2\mu_0}{2}\left(\vb{\pi}+\frac{k}{2}a\right)^2+V
\lb{Ha}
\ee
with the effective potential
\be
V=\frac{4\pi}{\mu_0}\left(a_ij^i+\frac{1}{c^2}a_0\rho^0\right)\ ,
\lb{Veff}
\ee
where $j^i=\epsilon^i_jJ^j$ and $\rho^0$ are some effective
current and charge distributions.

Making the identifications
\be
\ba{rcl}
     k&=&\displaystyle \frac{2\ e\ B}{\mu_0}\ ,\\[5mm]
\gamma^2&=&\displaystyle \frac{1}{c^2}\frac{\mu_0}{m_e}\ ,
\ea
\ee
where $e$ and $m_e$ stand for the physical charge and mass of the electron and
$\mu_0$ is the \textit{magnetic constant}.

The Hamiltonian is easily rewritten as
\be
H_B=\frac{1}{2m_e}\left(\vb{p}+e\ B\ a\right)^2+V\ .
\lb{HB}
\ee
It can be recognised as the Hamiltonian for
electrons on a potential under an external
perpendicular uniform magnetic field $B$.
The momenta $\vb{p}$ stands for the nonrelativistic
momenta and the original Chern-Simons coefficient $k$ is, in this way proportional
to the external magnetic field. This means
that the external magnetic field is related with
the topological character of the Quantum Field Theory.

The coupling constants on the original TQFT were
introduced in a not so standard
way to achieve the correct phenomenology of the model
which will be addressed in the last section of the paper.
Their units on SI are
\be
\ba{rcl}
[k]&=&A^2\ m^{-1}\ s\ ,\\[5mm]
[\gamma^2]&=&A^{-2}\ m^{-1}\ ,
\ea
\ee
With this convention the gauge field has units of
length $[a]=m$.

The external magnetic field $B$ and the
\textit{internal} one $b$ should not be confused.
The first one is encoded in the Chern-Simons coefficient
and is imposed externally to the system. The second one
is due to non trivial configurations of the
transverse gauge fields. Also it is important to stress that the
$b$ field is dimensionless, its dimensionfull counterpart
can be taken from the expression~(\ref{q}) for the
charges on the theory such that
\be
\tilde{b}=\frac{B}{4\pi} b\ .
\ee
The $2+1D$ magnetic field $\tilde{b}$
spreads over the thickness of the system on the
third spatial dimension, in here is assumed that
effectively it spreads uniformly over the
distance $\alpha_T$ and is null away from the
planar system where $\alpha_T$ stands for
the thickness of the system on the third spatial direction (perpendicular
to the $x\times y$ plane)

The vortex operator~(\ref{UV}) in this effective limit
is defined as
\be
V_\lambda=\exp\left\{\frac{i}{\hbar}\,\partial_i\lambda.\left(p^i+e\ B\ a^i\right)\right\}
\lb{Va}
\ee
up to a multiplicative constant due to the potential
which we can include in the normalization of the operator.

Concerning the nonrelativistic quantum problem,
the symmetric and Landau gauge correspond to
\be
\ba{rcl}
a_s&=&\displaystyle\left(-\frac{y}{2},\frac{x}{2}\right)\ ,\\[3mm]
a_L&=&\displaystyle\left(0,x\right)
\ea
\ee
respectively and both are related by a gauge
transformation $\lambda'=xy/2$, such that $a_s=a_L+d\lambda'$.

We considering the effective potential~\ref{Veff}
to be periodic with unit cell $\vb{dr}=(d_x,d_y)$.
The previous local $U(1)$ gauge transformations operators $V_\lambda$~(\ref{Va}) 
are therefore identified with lattice translations operators.
For the case of the symmetric gauge, with $\lambda=\vb{dr}.\vb{r}$,
they correspond to magnetic translation operators
while for the Landau gauge they correspond to the generators
of $U_q(sl_2)$ as will be computed in the next subsection. In this sense both the
Magnetic Translation Group and $U_q(sl_2)$ correspond to
distinct discrete subgroups of $U(1)$ with linear gauge parameter and are
related by a gauge transformation $\lambda'=xy/2$.

\setcounter{equation}{0}
\section{\lb{sec.AH}The Azbel-Hofstadter Model}

The Azbel-Hofstadter~\cite{AZ_0,HO_0} is a model that describes
Bloch electrons, i.e. electrons on a $2D$ periodic potential under a strong
perpendicular magnetic field. It is a model on the lattice and the
energy eigenvalues are computed by demanding the wave functions
solutions periodicity to be compatible with the lattice (see~\cite{wieg_1,wieg_2} for exact asymptotic
solutions). In this way one gets the band/gap structure
(depending on the magnetic flux per unit cell)
of the theory which constitutes a fractal
(Cantor set) known as the Hofstadter butterfly
(see~\cite{wieg_3} for a recent review).
A key point of this model is that the periodicity of wave functions
is only possible for rational magnetic flux per unit cell.
The physics for irrational values of the magnetic flux
can nevertheless be studied by
considering a succession of rational values
which converge to that irrational value. Even
in this irrational limit the structure of the
spectrum is maintained although the number of
bands and gaps become infinite.

\subsection{Periodic Potential and the Magnetic Translations Group}

Take now a periodic potential represented by a $2D$
lattice of unit cell with sides $d_x$ and $d_y$
(i.e. a planar rectangular crystal) of size $L_x$
and $L_y$ with periodic boundary conditions on both
directions $x$ and $y$. This means our space becomes
a square torus of area $A_{T^2}=L_xL_y$. For other
boundary conditions one would get other topologies.

In the presence of a periodic potential the spatial
translation symmetry is broken down to a discrete
version, which spans the equipotential points on
both directions of the potential lattice.
When a Magnetic field is present the theory is no longer
symmetric under the usual translations, but it still is under
the magnetic translations.
On the presence of a periodic potential
and for a particular gauge choice (Landau gauge),
the magnetic translations are reduced to only two generators
that constitute the $U_q(sl_2)$ quantum group.

To see it explicitly let us built the magnetic generators~\cite{brown}
\be
T(\vb{R}_d)=\exp\left\{-\frac{i}{\hbar}\vb{R}_d.(\vb{\pi}-e\ B a)\right\}
\ee
such that $\vb{R}_d$ stands for some vector of the potential lattice.
Note that these operators indeed commute with the Hamiltonian
\be
[T(\vb{R}_d),H_B]=0
\ee
since the potential value is the same at points related by the
shift $\vb{R}_d$ and the commutation relation $[\vb{\pi}-e\,B\,a,\vb{\pi}+e\,B\,a]=0$
holds.
Note that in the absence of external magnetic field these
operators become simply the translation operators
$T(\vb{R}_d)=\exp\{-i\vb{R}_d.\vb{\pi}/\hbar\}$ as argued
before.

Furthermore only two of such generators are necessary to span each
of the equipotential spaces. They are
\be
\ba{rcl}
T_x&=&T(\vb{R}_{d_x})\\[4mm]
T_y&=&T(\vb{R}_{d_y})
\ea
\ee
They indeed work as local gauge operators $V_a$ given by~(\ref{Va})
for the symmetric gauge with parameters
$\lambda_x=d_x x$ and $\lambda_y=d_y y$.

Acting on wave functions we get
\be
\ba{rl}
T_x\Psi(x,y)=&\displaystyle\exp\left\{i\,\frac{e\ B}{\hbar}d_xd_y\right\}\Psi(x+dx,y)\\[4mm]
T_y\Psi(x,y)=&\displaystyle\exp\left\{i\,\frac{e\ B}{\hbar}d_xd_y\right\}\Psi(x,y+d_y)\\[4mm]
\ea
\ee
in accordance to~(\ref{UPsi}).

Also one gets
\be
T_xT_y=\exp\left\{-i\,\frac{e\ B}{\hbar}d_xd_y\right\}T_yT_x
\ee
in accordance with~(\ref{VVcom}).

To complete the definition of the group consider
\be
T_xT_y=\exp\left\{-i\frac{e\ B}{2\hbar}d_xd_y\right\}T_{x+y}
\ee

So we just defined the group of magnetic translations
and become apparent that it is described
correctly in terms of discretized local $U(1)$ gauge operators
of a compact Abelian Maxwell Chern-Simons Massive Gauge
theory. Moreover in the Topological Quantum Field Theory
they are vortex operators. Here, in the Azbel-Hofstadter model,
they are interpreted as the generators
of the quantum group that generalises translations
on the presence of an external magnetic field.
This group can easily be interpreted as the possible
interpolations (tunnelling effect) between
non-distinguishable field
configurations of the theory or equivalent vacua,
also commonly called on the $3D$ Field Theory framework
monopole-instantons.

\subsection{Harper Equation from Landau Gauge}

For the Landau gauge $a_L=(0,x)$ the previous group
generators acting on wave functions are
\be
\ba{rcl}
\displaystyle T_x\Psi(x,y)&=&\displaystyle \exp\left(i k_x\right)\Psi(x+dx,y)\\[4mm]
\displaystyle T_y\Psi(x,y)&=&\displaystyle \exp\left(i k_y\right)\exp\left\{i\pi\omega\right\}\Psi(x,y)
\ea
\ee
where $\omega$ is the flux per unit cell over $\hbar$
\be
\omega=\frac{eB}{\hbar}d_xd_y
\ee
In this gauge the group has a one-dimensional representation
for each Bloch wave vector $(k_x,k_y)$.
There is no shift on the $y$ direction because
in the Landau gauge the canonical momentum $p_y$
is null (see the original works~\cite{H,brown,AZ_0,HO_0}
for further details).

The theory is in this way mapped into a one
dimensional model by writing the Hamiltonian as
\be
H=T_x+T_x^{-1}+\lambda\left(T_y+T_y^{-1}\right)
\ee
where $\lambda=d_y/d_x$. The resulting discretized shr\"{o}dinger equation
is the well know second order Harper equation~\cite{H}
\be
\psi_{n+1}+\psi_{n-1}+2\lambda\cos(2\pi(\omega n+\phi_0))\psi_{n}=E\psi_n
\ee
where $k_x=0$, $\phi_0=k_y$ and $E$ is the energy
eigenvalue. Under the map $x_n=\psi_{n-1}/\psi_n$ the first order
Harper map is obtained~\cite{KS}
\be
\ba{rcl}
x_{n+1}&=&\displaystyle-\frac{1}{x_n-E+2\lambda\cos(2\pi\phi_n)}\\[4mm]
\phi_{n+1}&=&\displaystyle\omega n+\phi_0
\ea
\ee

The Lyapunov exponent corresponding to the $x$ dynamics is given by
\be
\ba{rcl}
\overline{y}&=&\displaystyle\lim_{N\to\infty}\frac{1}{N}\sum^{N}_{n=0}y_n\\[4mm]
y_n&=&\displaystyle\log\frac{\partial x_{n+1}}{\partial x_n}=\log x_{n+1}^2
\ea
\ee

Aubry and Andr\'e~\cite{aa} proved that the Lyapunov exponent
$\overline{y}=-2\gamma$, being $\gamma$ the localisation length and
that $\overline{y}\le 0$ always. For $\overline{y}=0$, the phase 
is Extended and for $\overline{y}<0$ the phase is
Localized.

\subsection{Phases of Azbel-Hofstadter model: Bands and Gaps}

Numerically the two phases of these maps where studied to some extend in~\cite{PC_0,PC_1,KS,aa}.
\begin{figure*}[ht]
\includegraphics[width=150mm]{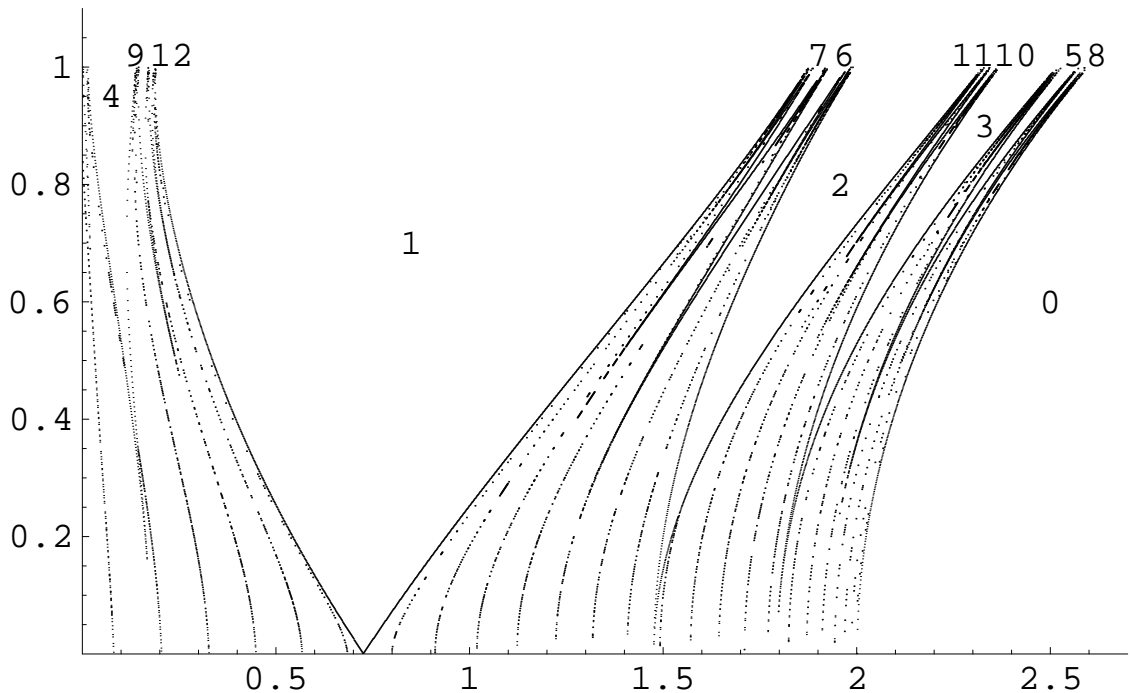}
\caption{\footnotesize Phase diagram for $\omega^*=(\sqrt{5}-1)/2$. The first 12
Localized phases are labelled in the diagram.}
\label{fig:phaseir}
\end{figure*}

Relating them with the original Azbel-Hofstadter one checks that
the Extended phases corresponds to the original energy bands and
the localized phases to the gaps.

In~\cite{PC_1} were studied the attractors in the Localized phases in the
$x\times\phi$ plane. It was suggested a classification of the many
localized phases of this map by integers $\eta$.
The magnetic flux $\omega^*=(\sqrt{5}-1)/2$ is taken to be
the limit of the succession $\omega_n=F_{n-1}/F_n$ where
$F_0=F_1=1$ and the relation $F_n=F_{n-1}+F_{n-2}$ define
the Fibonacci succession.

The full phase diagram for that irrational magnetic
flux $\omega=(\sqrt{5}-1)/2$
is shown in figure~\ref{fig:phaseir}.

\begin{figure*}[ht]
\includegraphics[width=150mm]{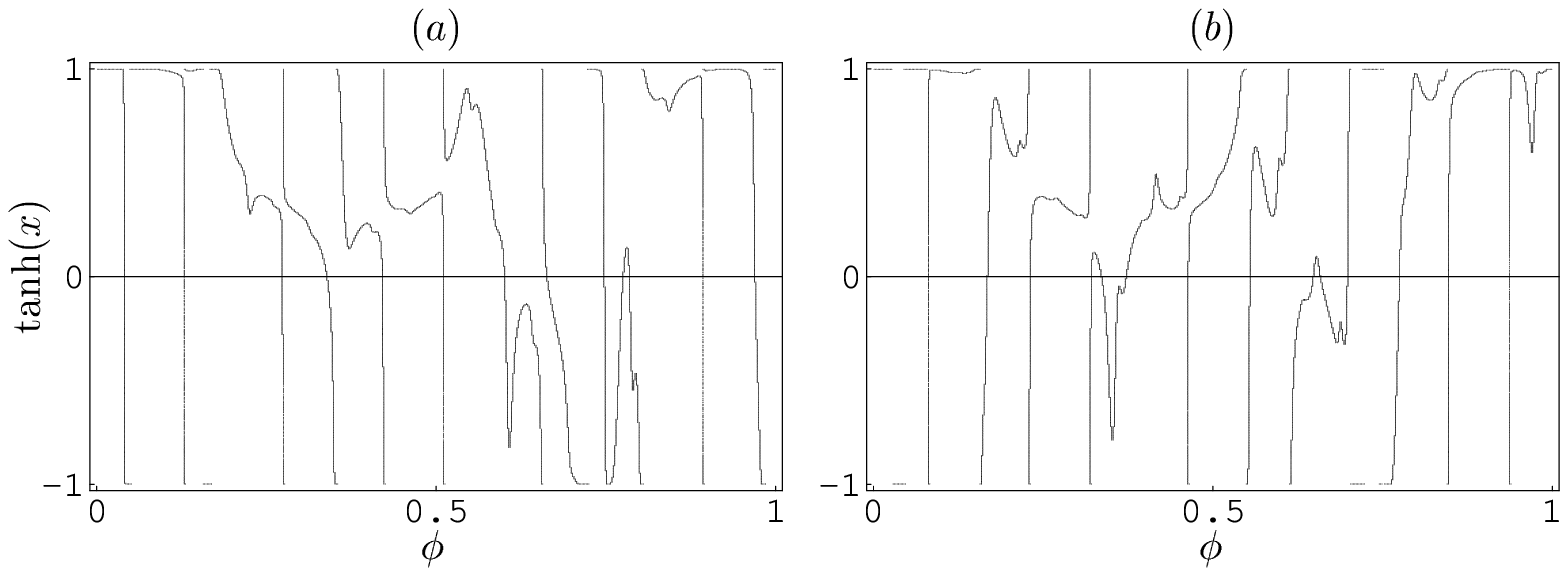}
\caption{Attractors for Localized phases on the torus $x\times\phi$
for $\omega^*=(\sqrt{5}-1)/2$ and $\lambda=1$.
(a) $\eta=11$, $E=2.342$; (b)$\eta=10$, $E=2.347$.}
\label{fig:wind}
\end{figure*}

The integers labels are the winding of the attractors on
the $x\times\phi$ torus. This torus is obtained by considering the identification
on $\phi:\ 0\cong 1$ and on $x:\ -\infty\cong\infty$ on the plane $x\times\phi$.
Figure~\ref{fig:wind} is an example of such attractors for $\eta=11$
and $\eta=10$.

Note that each winding corresponds to $x$
going trough $x=\pm\infty$ and crossing, necessarily, the line $x=0$. 
The first case means simply that the wave function is going trough zero and
does not constitute a problem, but the second one is
translated in terms of the original wave function into the existence
of divergences for the null values of $\phi$. These are essential singularities.
Therefore the conclusion is that the solutions of the wave
functions do exist on the Localized phases (gaps) but are not physical since they
are not normalizable.

Then the question which remains to answer is if the localized phases
and all the structure described in~\cite{PC_1} does correspond to
any physics at all or if it is just a mathematical
curiosity.

\setcounter{equation}{0}
\section{\lb{sec.gaps}Physics on the Gaps}

As explained so far one manage to go from a Topological Quantum Field Theory
to a quantum mechanics model of
nonrelativistic electrons on magnetic fields and then mapped it
into a simple dynamical system described by an one-dimensional map.

At the dynamical system level it was concluded that there are two kind
of phases, one Extended and the other one Localized. While on the
Extended phase the map have $2D$ attractors and the iterations
spread in the $x\times\phi$ plane according to some probability
distributions~\cite{PC_0}, on the Localized phases there are stable
one-dimensional attractors.

At the nonrelativistic quantum level these results must be interpreted in other
terms. The Azbel-Hofstadter model is precisely based on finding the
solutions of the Shr\"odinger equation which preserve the magnetic
translations symmetry. Therefore the previous Extended phases
correspond to the energy bands of the quantum system while the
Localized ones to the gaps of the theory. So,
although numerical solutions for the Schr\"odinger
equation were found, they are not physical since
are not square integrable due to \textit{essential}
divergences. Something must be missing in our picture.

\subsection{Symmetry Breaking and Phase Transitions}
 
Let us go back to the TQFT framework and
re-examine the previous results. The magnetic
translations are now the gauge transformations, the
Extended phases correspond to a regime where
the gauge symmetry is present while the Localized ones
to a regime where the gauge symmetry is somehow broken.

The issue here is symmetry breaking and the
theory being renormalizable or not.
In the broken regime the theory is nonrenormalizable
which translates to our nonrelativistic quantum limit as the
wave functions being non normalizable.
So the problem to solve is a very old one:
symmetry breaking maintaining the theory renormalizable!

As widely known the symmetry breaking must however be
\textit{nonexplicitly} in order that the broken
theory is still renormalizable. This issue has been well studied and is in the origin
of the Higgs-Kibble mechanism~\cite{Higgs_1,Higgs_2,Kibble}
of spontaneous symmetry breaking. There the symmetry
is broken by choosing a particular
point of the several degenerated vacua. 
For the setup presented here,
the natural way out, is to consider specific field
configurations allowed on the theory which break
the gauge invariance. Again this is an old subject,
we have some degenerate vacua and we are choosing
some particular configuration which minimises the
energy of the system constituting then a
\textit{classical} background.
These are often called
monopoles~\cite{thooft_1,thooft_2,polyakov_1,polyakov_2,thooft_3}
and constitute what is know as dynamical symmetry
breaking. By allowing such configurations the
gauge symmetry is broken in exchange of field
configurations but the theory is still renormalizable.
This is also the same mechanism which in M/string-theory
compactifications break the supersymmetry due to the
choice of some particular vacua manifold
(classical solution of the gravity theory).

Note that in $2+1$D the dynamical Symmetry breaking
is also associated to a confined-condensate phase
transition of the theory as will be discussed to
some extension later on.

\subsection{Monopoles and White Holes}

Consider now the results computed on the work of
Stichel~\cite{St_2} (see also its references and in
particular~\cite{St_0,St_1}) where a classical deformation of the gauge
field yields both a monopole-like configuration and a
deformation of the metric which constitutes a gravitational white
hole. Note that although the framework of that paper is different it
translates to the framework presented here if in the
decomposition~(\ref{Adef}) $\vb{A}=\vb{a}_1+\vb{a}_2$ the second field
$\vb{a}_2$ is considered at classical level to be a deformation of the
first one $\vb{a}_1$.

There the original theory is of a nonrelativistic point particles coupled
to gauge fields which are deformed in order to allow time-dependent
area preserving diffeomorphisms. Here the starting point is a quantum
field theory which has as a particular nonrelativistic
quantum mechanical limit, classical electrons
travelling on a perpendicular magnetic field.
Some non-trivial field
configurations on the original quantum field theory will translate, in
the classical limit, to a deformation of the gauge field and a deformation
of the metric which constitutes a geometric bag or monopole.
It is this deformation that
interests us in the scope of this work. Also note that as explained before,
at the TQFT level, any magnetic field depends only on the field $\vb{a}_2$.
The existence of non trivial magnetic field configurations
breaks the gauge symmetry since the path integral is not invariant
under large gauge transformations (see for instance~\cite{AHPS}),
this is also the reason why in~\cite{AFC_1} there are only
functional wave functions for trivial magnetic field configurations
$\int b=0$.

Physical particles are called in $2+1D$ vortexes while
the term monopoles is associated with instantons
(tunnelling processes). In this work the term
magnetic monopole is used to denote physical
\textit{static} particles constituted by gauge field
configurations on the $2D$ spatial plane.

It is important to stress that the configurations presented
here are pure magnetic monopoles with finite energy. This
is only possible due to the inclusion of the
nontrivial gravitational configurations.

Let a generic gauge transformation be decomposed into two components
\be
\Lambda(\vb{x},t)=\sum_\alpha\phi(\vb{x}-\vb{x}_\alpha)+\lambda(\vb{x},t)\ ,
\ee
where $\lambda$ is a regular function and $\phi$ is a spatial singular
function
\be
\phi(\vb{x})=\arctan\frac{x}{y}\ .
\ee
Take only one singular point at $(x,y)=(0,0)$ and consider 
a time independent \textit{hedgehog} solution
for the gauge fields $A$
\be
\ba{rcl}
& & \\
A_x&=&\displaystyle y\left(1-\sqrt{1-\tilde{\theta}/r^2}\right)\frac{2N}{\tilde{\theta}}\ ,\\[6mm]
A_y&=&\displaystyle-x\left(1-\sqrt{1-\tilde{\theta}/r^2}\right)\frac{2N}{\tilde{\theta}}\ ,\\[6mm]
A_0&=&\mathrm{const}\\[4mm]
\ea
\lb{Asol}
\ee
which constitutes a monopole configuration,
and the corresponding metric solution
\be
\ba{rcl}
& & \\[2mm]
h^{xx}&=&\displaystyle\frac{1}{\sqrt{2}}\left(\frac{1}{1-\frac{\tilde{\theta}}{r^2}}-\frac{\tilde{\theta}(2-\frac{r^2}{\tilde{\theta}})}{r^2-\tilde{\theta}}\frac{x^2}{r^2}\right)\ ,\\[6mm]
h^{xy}&=&\displaystyle-\frac{1}{\sqrt{2}}\frac{\tilde{\theta}(2-\frac{r^2}{\tilde{\theta}})}{r^2-\tilde{\theta}}\frac{xy}{r^2}\ ,\\[6mm]
h^{yy}&=&\displaystyle\frac{1}{\sqrt{2}}\left(\frac{1}{1-\frac{\tilde{\theta}}{r^2}}-\frac{\tilde{\theta}(2-\frac{r^2}{\tilde{\theta}})}{r^2-\tilde{\theta}}\frac{y^2}{r^2}\right)\ ,\\[6mm]
\ea
\lb{metric}
\ee
where $r^2=x^2+y^2$ stands for the planar radius and $N$ is an integer with units of
length square $[N]=m^2$ that parameterises the monopole charge.

The parameter of these configurations is given by
\be
\tilde{\theta}=e^2\alpha_n^4\frac{k}{4\pi c}=\frac{e^7}{2\pi}\left(\frac{\mu_0}{m_e}\right)^4\frac{B}{c\mu_0},
\ee
with units of length square $[\tilde{\theta}]=m^2$.
It is therefore proportional to the external perpendicular magnetic field 

The electric field is null since $A_0=\textrm{const}$
and the magnetic field of
the configuration is computed to be
\be
b(r)=\frac{2 N}{\tilde{\theta}}\left(2+\frac{\tilde{\theta}-2r^2}{r^2\sqrt{1-\frac{\tilde{\theta}}{r^2}}}\right)\ .
\ee

The determinant of the inverse metric is
\be
h^{-1}(r)=\frac{r^2}{r^2-\tilde{\theta}}
\ee
and as expected the metric is asymptotically flat
\be
\lim_{r\to\infty}h(r)=1\ .
\ee

The gravitational configuration corresponds in this nonrelativistic
quantum mechanical limit to a white hole with radius
\be
r_0^2=\frac{e^7}{2\pi}\left(\frac{\mu_0}{m_e}\right)^4\frac{B}{c\mu_0}
\ee
which reflects any incident wave function as described in detail in the last
section of~\cite{St_2}.

By request of the referee we proceed to analyse the properties of
the monopole-white hole and compute the asymptotic solutions of the Schr\"odinger equation. 

In planar polar coordinates the metric is given by
\be
ds^2=\sqrt{2}dr^2+\frac{r^4}{\sqrt{2}(r^2-\theta)}d\varphi\ .
\ee
Although the curvature is null everywhere, $R=0$ as expected in
two dimensional gravity, the contraction of two Ricci Tensors is
\be
R_{ij}R^{ij}=-\frac{(r^2-2\,\theta)^4}{r^4\,(r^2-\theta)^4}\ .
\ee
The vanishing $\theta$ limit is well defined, $\lim_{\theta\to 0}R_{ij}R^{ij}=1/r^4$
corresponding to the flat metric, as it should since $\theta$ is the deformation parameter.
Clearly $r=r_0=\sqrt{\theta}$ is singularity. We will next show that it also constitutes a
\textit{white-horizon} such that the wave function vanishes at $r=r_0$.
This fact agrees with the interpretation of the monopole-white hole being a physical particle of radius $r_0$.
We note that the deformation of the space induces a planar angle deficit, hence $r_0$ is a conical
singularity. The angular variable is defined only on the range $[0,2\pi\sqrt{1-\theta/r^2}[$.

To show that the wave functions are well defined in the
presence of these particle and they don't penetrate the core,
take the gauge field $\vb{a_2}$ as the deformation, such that
the monopole configuration is due to this field.
The final deformed Hamiltonian for the gauge field $\vb{a_1}$
is rewritten in the symmetric gauge as
\bea
m_eH_{def}&=&\displaystyle\left(\pi_i-eB\tilde{\epsilon}_{ik}x^k\right)\left(\pi^j-eB\tilde{\epsilon}_{jl}x^l\right)h^{ij}\nonumber\\[5mm]
       &=&\displaystyle\sqrt{2}\pi_r^2+\frac{(l-eBr^2)^2}{\sqrt{2}(r^2-\theta)}\ ,\hspace{20mm}
\eea
where $\tilde{\epsilon}^{ij}=\sqrt{h}\epsilon^{ij}$ is the covariant antisymmetric tensor,
the vacuum energy~(\ref{Emon}) is shifted, $(x^1,x^2)=(x,y)$ and
$l$ is the canonical angular momentum $l=\tilde{\epsilon}_{ij}\,x^i\,h^{jk}\,\pi_k=\sqrt{h}\,\pi_\varphi$.

In the following we compute the wave function solution using the Schr\"odinger picture
with $\pi_r=\hbar\partial_r$ and Moyal-Weyl quantization procedure as described in~\cite{St_2}. Basically the
normal ordering prescription for a product of operators corresponds to all possible monomial combinations,
or equivalently stated we consider the full (noncommutative) star product of operators $f_1(\hat{p})\star f_2(\hat{q})=\exp\{-i\hbar(\partial^2/\partial p\partial q)\}f_1(p)f_2(q)|_{p\to\hat{p},q\to\hat{q}}$
(see eq~(3.4-3.7) of~\cite{castellani}), which in our simple Hamiltonian is exactly an expansion
to order 2 on $\hbar$. Therefore we obtain the Scr\"odinger equation
\bea
\left[E-\frac{(l-eBr^2)^2}{\sqrt{2}m_e(r^2-\theta)}+\right.\hspace{40mm}\\[2mm]
\left.\frac{\sqrt{2}\hbar^2}{m_e}\left(\frac{1}{\sqrt{h}}\partial_r\sqrt{h}\partial_r+\frac{r^4(3r^2+\theta)}{\sqrt{2}(r^2-\theta)^4}\right)\right]\psi=0\nonumber
\eea

Taking the limit of $r\to\sqrt{\theta}$, we solve the asymptotic sch\"odinger equation, finding
\begin{widetext}
\be
\psi(r)=(r^2-\theta)^\frac{25}{8}(r^2-2\,\theta)^{-\frac{7}{2}+\frac{(l-2\,e\,B\theta)^2}{8\,\hbar^2}}
\exp\left\{\frac{1}{16}\left(\frac{e\,B\,r^2\,(4\,l-e\,B(r^2+4\theta))}{\hbar^2}-\frac{(30\,r^2+26\,\theta)\theta}{(r^2-\theta)^2}\right)\right\}\ .
\ee
\end{widetext}
In the limit $r=\sqrt{\theta}$ the wave function vanishes independently of the angular momentum eigenvalue $l$. Note
also that the asymptotic solution does not depend on the energy of the solution. Then we proved that effectively
the electron wave function does not penetrate the core of the monopole-white hole.

One can now interpret the white hole, a gravitational deformation
induced by the gauge fields, as being a physical particle,
the monopole or a geometric bag. A full description would also need
to take into account the white hole quantum numbers. This issue
is not going to be studied here, see for instance~\cite{Ply} (and references therein).
In the scope of this work, let us simply compute some relevant
quantities of such a particle, the energy, charge and mass density.

The energy of the configuration is finite as required 
and is just the integral of the magnetic field squared
\be
\ba{rcl}
E_{mon}&=&\displaystyle\frac{\pi\ \mu_0\ \alpha_T}{4\ \gamma^2\ V_3}\int dr\ h^{-1}(r)\ r\ b(r)^2\nonumber\\[4mm]
&=&\displaystyle\frac{N\ \pi\ \mu_0\ \alpha_T}{4\ \gamma^2\ V_3}\ \left(3-\ln 2\right)
\ea
\lb{Emon}
\ee

The physically measurable charge of the magnetic monopole is
\be
Q_m=\frac{B}{4}\int dr\ r\ b(r)=\frac{B}{4}N
\lb{Qm}
\ee
as expected. For $N=1$ we have the fundamental charge allowed on the theory.

Considering that the monopole is spherical (in $3D$ terms),
the constant $\alpha_T$ is of the same order of magnitude of the white
hole's radius. Therefore we obtain
\be
\rho_{mon}=1.195\ 10^{17}\frac1{B}\ .
\ee
For strong values of the magnetic field (say of order
$B^2=25\ T^2$ - see equation~(\ref{Bstrong}) in the end of the article)
we get a value for the density of order $10^{16}\ Kg m^{-3}$. This is
somewhere between the density of the electron ($10^{13}$) and nuclear
matter ($10^{18}$ - considering $r_p\sim 1\,fm$).

As a final remark note that the procedure described in~\cite{St_2}
takes in account the gauss law constraint and therefore
gauge fixes the theory. The point is that on a theory containing gravity
both the degrees of freedom of gravity and the gauge theory will
be constrained together such that for a particular nontrivial field
configuration the gauge symmetry is broken (or gauge fixed if one
prefers to say so) in exchange of a nontrivial gravitational
background. Also it must be stressed that the metric~(\ref{metric})
is not the one presented on~\cite{St_2},
there the ansatz corresponds to an anion (with both electric
and magnetic charge) and renders the determinant of the metric
to be always $1$. The metric presented here has determinant
$1$ only asymptotically rendering finite
energy pure magnetic monopole-white hole configurations.
In terms of the quantum field theory these issues
must be translated into the existence of a gravitational
sector of the theory. This is not going to be studied here.

\subsection{Confinement and Condensation of Monopoles}

Let us analyse the results of the previous subsection at the
level of the TQFT.
In the early works on Topological Massive Gauge Theories (see for
instance~\cite{AHPS}) the monopole charge was imposed to average to
zero over the entire manifold in order to preserve gauge invariance of the
path integral.
It was studied in~\cite{AK_5} that $2+1$D
compact Maxwell Chern-Simons has two distinct phases. In the confined
phase the magnetic charge is confined such that the monopoles are
paired. The total charge is null for each pair and the monopole
charge is in this way screened. This is the phase in which gauge
symmetry is unbroken, there are only trivial configurations of magnetic
field. In the condensate phase the magnetic charge condensates
and the monopoles are not paired any longer. The
total magnetic charge has now a non null value~\cite{AK_6,AK_7}.
See figure~\ref{fig:mon} for a suggestive picture of both phases.
\begin{figure*}
\includegraphics[width=100mm]{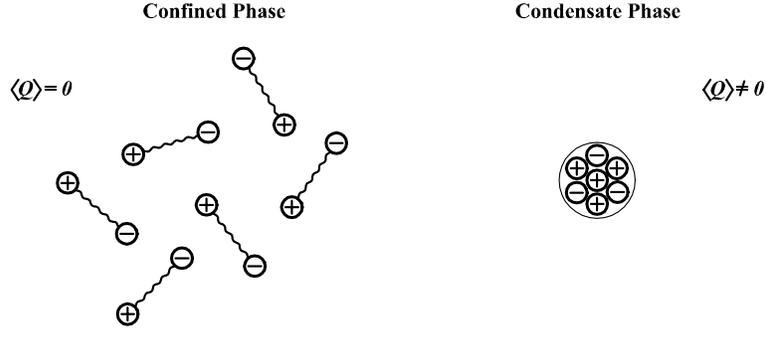}
\caption{\lb{fig:mon}Monopole confinement and condensate.}
\end{figure*}
The above arguments strongly suggest that the existence of
monopole condensation on the TQFT is intrinsically linked to
the deformation of the classical gauge field and metric in
the classical limit.

Furthermore it was argued in~\cite{AK_6} that in the continuum
limit of the Abelian theory the vortexes are not present due to being
logarithmically UV divergent and therefore the monopole effect would
be negligible anyway. The mechanisms which regularize the theory
seem to be two folded. Both considering a periodic current in the
same manner that the periodic potential is considered in the original Azbel-Hofstadter
model such that the theory is always defined on a lattice and taking
in account gravitational effects in order to correctly regularize the
theory. Given the framework of the Azbel-Hofstadter
model the theory is defined on a lattice due to the periodicity of the
external current (the effective periodic potential). This is not enough
since the wave functions are not normalizable and one has to account
also for gravitational configurations.

\subsection{Divergences Removed on the Gaps, Chains of Filaments and Monopoles}

The punch line is that, if we allow configurations where monopoles have a
non null net charge together with a gravitational deformation,
they will preserve the finitude of the wave function.
Moreover the value of the net charge must still be a
multiple of the fundamental magnetic charge due to the fact that the
full effect is still due to a set of such particles. The
monopole number would be related to the number of essential (nonremovable)
singularities of the bare wave function which are present in the
pure gauge theory. That number is simply given by the winding number of the dynamical system
on the $x\times\phi$ plane, that is the classification index of the phase
diagrams on the last section!

Furthermore if one takes the wave function to be defined as a function
of the phase $\phi$ one knows that the divergences are located on
the points where the attractors are null. These will be the points
where the white holes must stand in order the theory to be well defined
and the wave functions divergences absent.

Translating it in more formal terms note that the phase
$\phi_n=\phi_0+\omega n$ parameterizes a discretized
line. For irrational values of $\omega$ we can define
a continuum phase $\phi$ which parameterises the attractors
on the $x\times\phi$ plane
\be
x(\phi)=\psi(\phi)/\psi(\phi+\delta\phi)
\ee
($x$ should not be confused with the spatial coordinate - see
figure~\ref{fig:wind}). For a given string of length $L_l$
the phase is simply a normalized parameter
\be
\phi=\frac{l}{L_l}\ \ \ l\in[0,L_l]
\ee

Now one can consider wave functions of the form
\be
\Psi(l)\sim e^{\int f(l).dl}
\ee
such that the function $f(l)$ defines the linear attractor on the localized
phases
\be
x(l)=\frac{\psi(l)}{\psi(l+\delta l)}=e^{F(l)-F(l+\delta l)}
\ee
where $F(l)$ stands for the value of the primitive of $f$ evaluated at $l$.

Once one considers the theory with monopole-white holes the wave function
is physically well defined and one obtains a chain constitute by electron filaments
attached to monopoles. The position of the monopole-white holes
on the chain will be located at the divergences of the bare wave functions
and its number will correspond to the before mention $\eta$ (the label of the phase).
See figure~\ref{fig:chain} for a picture of such a chain.
\begin{figure}
\includegraphics[width=75mm]{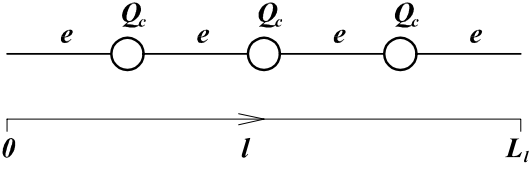}
\caption{\lb{fig:chain}Chain containing 3 monopole-white holes of fundamental
charge $Q=k/4$ and electron filaments $e$.}
\end{figure}

\subsection{Possible Phenomenology}

A experimental analysis is out of the scope of this work and the expertise of the author.
However a brief phenomenological analysis is carried out for the sake of completeness.

To have an order of magnitude of the physical effects
consider some \textit{strong} magnetic field $B$, meaning that
\be
\mu\,B>>1
\lb{Bstrong}
\ee
where $\mu$ is the electron mobility. Taking it to be of order
$\mu\sim 10 m^2V^{-1}s^{-1}$~\cite{helium} one requires a magnetic field of order
$B\sim 5\ T$.
Then the radius of the monopole-white hole is of order
\be
r_0=\sqrt{\frac{e^7}{2\pi}\left(\frac{\mu_0}{m_e}\right)^4\frac{B}{c\mu_0}}\sim 10^{-15} m
\ee

This results indicate that in principle it is possible to test the theoretical results
presented in this work with present day technology.

\begin{acknowledgements}
This work was supported by SFRH/BPD/5638/2001 and SFRH/BPD/17683/2004.
The author thanks everyone who contributed to the development of the ideas presented here,
helped with several technical difficulties and pointed out literature.
\end{acknowledgements}

\end{document}